# Molecular motors govern liquid-like ordering and fusion dynamics of bacterial colonies

Anton Welker*, Tom Cronenberg*, Robert Zöllner*, Claudia Meel, Katja Siewering, Niklas Bender, Marc Hennes, Enno R. Oldewurtel, Berenike Maier

* equal contribution

Institute for Biological Physics, University of Cologne, Zülpicher Str. 77, 50539 Köln, Germany

Bacteria can adjust the structure of colonies and biofilms to enhance their survival rate under external stress. Here, we explore the link between bacterial interaction forces and colony structure. We show that the activity of extracellular pilus motors enhances local ordering and accelerates fusion dynamics of bacterial colonies. The radial distribution function of mature colonies shows local fluid-like order. The degree and dynamics of ordering are dependent on motor activity. At a larger scale, the fusion dynamics of two colonies shows liquid-like behavior whereby motor activity strongly affects tension and viscosity.

Cocci are spherically shaped bacterial cells. Often, they produce surface structures that mediate attractive interactions between bacteria. Spheres with attractive interactions are reminiscent of non-living colloidal systems with attractive interactions that tend to form liquid – or crystal-like structures. Assembly of colloidal crystals can be well controlled for example by DNA-hybridization or through electrostatic interactions [1-3]. Thereby dynamics and efficiency of equilibration are very sensitive to the form and range of interaction potentials which are difficult to control experimentally [1,4].

In contrast to non-living colloidal systems, bacterial cocci reproduce. During reproduction, their shape changes from spherical to dumbbell-shaped. Another important difference to non-living colloids is the use of attractive surface appendages with molecular motor properties. It is currently unclear how motor activity affects the structure of bacterial colonies. Here, *Neisseria gonorrhoeae* (gonococcus) and its type IV pilus (T4P) motors serve as a model system for addressing these questions. The gonococcus has a spherical cell body with a diameter of ~ 0.7 µm. The extracellular T4P polymers are randomly located across the cell surface [5,6]. T4P act as depolymerization motors; as T4P polymers depolymerize they retract and this retraction process can generate forces exceeding $F > 100$ pN onto a load attached to the pilus [7]. Pilus retraction is powered by the retraction ATPase PilT located at the base of the T4P machine (Fig. S1) [5]. When individual cells are attached to surfaces, multiple T4P cooperate for persistent movement through a tug-of-war mechanism [8,9]. When piliated bacteria are in close proximity to each other, T4P – T4P interactions induce self-assembly into colonies [10]. Self-assembly is reversible in young colonies [11]. Because deletion of T4P reduces the attractive interactions below the detection limit for gonococci [12,13], mutant strains in T4P-related genes provide a tool-box for tuning bacteria-bacteria interaction forces.

In this study, we test the hypothesis that spherical bacteria behave like a hard-sphere liquid with genetically tunable interactions. In our system, attractive interactions between bacteria are governed by the T4P motor. We show that molecular motor activity accelerates local liquid-like ordering and shape relaxations during colony fusion. Taken together, our results directly link molecular and macroscopic parameters by correlating inter-bacterial forces and binding probabilities with the spatio-temporal dynamics and materials properties of bacterial colonies.

*Colonies show liquid-like order.* - First, we investigated whether bacteria showed short-range order within colonies. To this end, wt* (Table S1) bacteria were incubated within a flow chamber under continuous nutrient supply for 6 h and 24 h, respectively. Subsequently, the colony structures were imaged using confocal microscopy. The dominant morphologies were spherical colonies (Fig. S2). A particle tracker was used to determine the three-dimensional coordinates of the cell bodies within the colonies (Fig. 1, Fig. S3). The bacteria were tightly packed within the colonies and the average density (number of nearest neighbors *n* within 1.4 µm of reference bacterium) tended to increase between 6 h and 24 h of colony growth (Fig. 1b,c, Fig. S3d). Within some colonies, holes were present after 24 h. Therefore, we calculated the remoteness [14], i.e. the probability of having a distance *r* from each voxel to the centroid of the closest bacterium (Fig. S3c). We found that the remoteness shifted to higher values after 24 h, in agreement with having more and larger areas that were not occupied with bacteria within the colony.

To address the question whether bacteria are ordered within the colonies, we determined the radial distribution function *g(r)*. *g(r)* is defined so that $N/V\, g(r) r^2 dr$ is the probability that the center of a bacterium will be found in the range *dr* at a distance *r* from the center of another bacterium, where $N/V$ is the number density of bacteria. At 6 h, *g(r)* showed two maxima (Fig. 1d). At higher cell-to-cell distances, the distribution is flat indicating that no order existed at distance exceeding r = 2.5 µm or three cellular diameters. After 24 h, the position of the first maximum shifted to a lower value. A lower average contact distances is in agreement with increased cell density and agrees well with Fig. S3d and Figs. 1 b, c showing an increase in the number of nearest neighbor cells. Moreover, the peaks became more pronounced indicating stronger ordering. The shapes of *g(r)* show good qualitative agreement with individual-based simulations that explicitly take T4P dynamics into account [15]. They are also reminiscent of the radial distribution functions found in colloidal systems with Lennard-Jones-like interactions [16]. Various analytical expressions for *g(r)* have been proposed for Lennard-Jones fluids and shown to fit the radial distribution function obtained from Monte-Carlo simulations very well [17,18]. Here, we used the formula proposed by Matteoli and Mansoori [17]

$$g(r > r_0) = 1 + y^{-m}[g(r_0) - 1 - \lambda] + \left(\frac{y-1+\lambda}{y}\right) exp[-\alpha(y-1)]cos[\beta(y-1)] \quad (eq.\,1)$$

where $r_0$ is the contact distance between two bacteria, $y = r/r_0$ and *m, λ, α,* and *β* are adjustable parameters (Table S2). In particular, the contact distance was found to decrease from $r_0^{6h} = (1.18 \pm 0.02)$ µm to $r_0^{24h} = (1.14 \pm 0.01)$ µm and the amplitude of the first maximum increased from $g(r_0)^{6h} = 1.23 \pm 0.02$ to $g(r_0)^{24h} = 1.46 \pm 0.02$. To summarize, gonococcal colonies show increasing local ordering and density with time.

*Non-functional T4P retraction ATPases affect pilus motor activity.* - Attractive force between gonococci is primarily generated by type IV pili (T4P) [12] [11]. Deletion of these hair-like appendages inhibits the formation of colonies. Here, we addressed the question whether local ordering required motor activity of the T4P. The idea is that T4P form a network between the bacteria and retraction of individual T4P generates a tug-of-war between bacteria (Fig. 1e), causing local order. To scrutinize this idea, we used a mutant strain lacking the T4P retraction ATPase PilT. These bacteria generate pili capable of supporting colony formation. However, pili cannot retract and therefore active force-generation is inhibited in this strain [19]. For tuning motor activities, we additionally generated strains $pilT_{WB1}$ and $pilT_{WB2}$ where we introduced non-functional $pilT_{WB}$ under the control of different promoters in addition to functional $pilT$. The rationale behind this construction was that non-functional $PilT_{WB}$ is likely to exert a dominant-negative effect on T4P retraction because PilT functions as hexamers that induce T4P retraction upon binding to the T4P complex [20]. Using a bacteria-two hybrid approach, we verified that the interaction between PilT-PilT and PilT-$PilT_{WB}$ were comparable (FIG. S6), strongly suggesting that PilT-$PilT_{WB}$ heterohexamers form.

We used our previously established T4P retraction assay for quantifying the effect of $pilT_{WB}$ expression on motor activity (FIG. 2a) [8]. The distribution of T4P retraction velocities shifted to lower values in strain $pilT_{WB1}$ compared to wt* and to even lower values for strain $pilT_{WB2}$ (FIG. 2b). In the presence of non-functional $PilT_{WB}$, the probability of finding a T4P in the state of elongation or pausing increased in strains $pilT_{WB1}$ and $pilT_{WB2}$ (FIG. S7). In conclusion, expression of non-functional $pilT_{WB}$ ATPases strongly affects T4P motor activity, namely motor velocity and the probability of finding the motor in the states of retraction, elongation, and pausing, respectively.

*Pilus motor activity affects bacterial interactions.* – We aimed at correlating changes in motor activity with changes of the interactions between bacteria. We used a dual laser trap setup for characterizing the dynamics of pilus-mediated interaction forces between two bacteria (FIG. 3a). A spherical bacterium was trapped within each of the laser traps. When the pili from different bacteria bind to each other and at least one of them retracts, the bacteria are attracted towards each other (FIG. 3b). We observed different states of pilus-pilus interactions including T4P retraction pulling both bacteria towards each other, elongation, where the bacteria moved away from each other, and pausing where *d* was constant but nonzero. Often, bacteria moved away from each other at a rate larger than 10 µm/s, indicating that the bond between pili had ruptured. Sometimes, the cell did not return to the center of the laser trap after a rupture event. In this case (denoted as bundling), we assume that more than one pilus was bound and while the shortest bond had ruptured, the second shortest bond kept the bacteria deflected from the center of the trap. The mean rupture force was $F^{wt*}_{rupture} = (50 \pm 24)$ pN for wt* (Fig. 3c). Neither strain $pilT_{WB1}$ nor strain $pilT_{WB2}$ showed significantly different distributions (KS-test) from the wt* distribution. Next, we investigated the frequencies at which the transitions to a specific state occurred (Fig. 3d). The frequency at which T4P retracted decreased in the presence of non-functional $PilT_{WB}$ while the frequency of elongations increased. Importantly, the rupture frequency decreased with in the presence of $PilT_{WB}$. The probability that two cells in the laser traps were connected through pili increased in strain $pilT_{WB1}$ and was highest in strain $pilT_{WB2}$ (FIG. 3e). This observation can be explained by the effect of $PilT_{WB}$ on motor activity. T4P tend to elongate rather than detach (FIG. S7) favoring T4P-T4P binding. Additionally, since the detachment rate is force-dependent [8] and the velocity of T4P retraction is lower (FIG. 2b), the probability of finding T4P attached to each other increases. Taken together, the newly designed strains allow us to tune the dynamics of pilus-mediated attraction through T4P retraction and the frequencies at which pili detach and thus release the interaction between adjacent cells.

*Active motors accelerate local ordering.* - The structures of the colonies formed by strains $pilT_{WB1}$ and $pilT_{WB2}$ were very similar to the wt* structure (FIG. 4a, b, FIG. S8). Interestingly, after 6 h the contact distances of these mutant strains were larger compared to the contact distance of the wt* (FIG. 4e, f). After 24 h, both the contact distance $r_0^{WB2,24h} = (1.13 \pm 0.01)$ μm and the amplitude of the first maximum $g(r_0)^{WB2,24h} = 1.48 \pm 0.02$ were comparable to the wt* parameters. When T4P retraction was fully inhibited by deleting *pilT*, both the local and the global colony structure were very different from the wt* structure (FIG. 4c, d, FIG. S9). The shape of microcolonies formed by *ΔpilT* cells was non-spherical in agreement with previous studies [10,21]. The cell density was lower compared to wt* colonies (FIG. S9). After 6 h of growth, *g(r)* showed no evidence for local order (FIG. 4d) indicating that active T4P retraction accelerated the process of local liquid-like ordering. After 24 h, a maximum became discernible, although its amplitude was lower compared to the wt* amplitude with $r_0^{\Delta pilT,24h} = (1.08 \pm 0.04)$ μm and $g(r_0)^{\Delta pilT,24h} = 1.30 \pm 0.02$. In summary, active force generation by T4P is not essential for local liquid-like ordering, but it enhances the rate at which local order and dense cellular packaging are achieved.

*Motor activity governs shape relaxations.* - We addressed the question whether T4P-mediated interaction forces supported liquid-like behavior of bacterial colonies at length scales beyond a few bacterial diameters. Like liquids, populations of self-attracting cocci form spherical colonies that coalesce upon contact [10,11]. We tested whether the dynamics of colony fusion were in agreement with the shape changes exhibited by an initially ellipsoidal liquid drop as it minimizes its surface area. Using the mutant strains expressing non-functional $pilT_{WB}$, we can tune the molecular interactions between the bacteria and investigate their effect on the macroscopic properties. During the fusion process, we assume that the energy change due to the reduction of surface area is balanced by the work of viscous deformation. We used the following fluid model of fusion dynamics to describe the shape changes during the fusion of two colonies [22,23]. Let the initial shape of the fusing colony be described by an oblate ellipsoid with the ratio of the minor axis $b$ and major axis $a$ $f = b/a$ and its volume $v = 4/3\,\pi ab^2$. Then the rate of deformation of the drop is

$$\frac{df}{dt} = \frac{\sigma}{v^{1/3}\eta}\rho(f) \quad (eq.\,2)$$

where $\sigma$ is the surface tension and $\eta$ the viscosity. $\rho(f)$ is described in the Supplementary Information. If at time $t_0$ an ellipsoid of cells has an axial ratio $f_0$ and by time $t$ it reaches $f$ then

$$v^{\frac{1}{3}}[\tau(f_1) - \tau(f_0)] = \frac{\sigma}{\eta}[t_1 - t_0] \quad (eq.\,3)$$

$\tau$ can be obtained by numerical integration of $\rho(f)$.

Experiments were initiated by inducing disassembly of colonies by oxygen depletion as described before [11]. Subsequently, oxygen-rich medium was applied triggering the formation of motile colonies (FIG. 5a). Once two colonies converged, they fused into a single colony whose projection in the xy-plane took the shape of an ellipse. FIG. 5b shows that $v^{1/3}[\tau(f) - \tau(f_0)]$ increases linearly with time in agreement with the dynamics predicted for the fusion of two highly viscous liquid drops (FIG. 5b). The dynamics of shape changes was slower for the $pilT_{WB1}$ strain compared to the wt* strain and the distribution of the ratio between surface tension and viscosity, $\sigma/\eta$, was shifted to lower values (FIG. 5c). For describing the short term dynamics of fusion, we used a different model describing the "neck" length of two fusing colonies as a function of time [24] (FIG. S10). At short time scales on the order of 10 s, the shapes relax faster than the model model predicts. At longer time scales the relaxation dynamics is well-described by the model and $\sigma/\eta$ agrees with FIG. 5c. We propose that the relaxation dynamics at short times is dominated by a layer of motile cells residing at colony surfaces [12].

Next, we relate the materials property $\sigma/\eta$ to the molecular interactions characterized in FIG. 3. Taking into account the rupture force $F_{rupture}$, the average number of pili per cell [6], and the probability that two cells are bound $p_{bound}$ through pilus-pilus interaction, we estimate the surface tension $\sigma$ (Supplementary Information) and obtain $\sigma_{wt*} \approx 5 \cdot 10^{-5}$ Nm$^{-1}$ and $\sigma_{WB1} \approx 8 \cdot 10^{-5}$ Nm$^{-1}$. Experimentally, we confirmed the right order of magnitude by deforming colonies through centrifugation (FIG. S11). Using these values, our experimentally determined ratios of $\sigma/\eta$ (FIG. 5) yield the viscosities of $\eta_{wt*} \approx 350$ Nsm$^{-2}$ and $\eta_{WB1} \approx 2000$ Nsm$^{-2}$. The increased probability of pilus-pilus binding in the *pilT$_{WB1}$* strain is consistent with a lower off-rate and explains increased viscosity. Using computer simulations, Pönisch et al showed that the shape relaxation strongly depend on T4P-T4P rupture force and detachment times [15]. Using rupture forces comparable to our experimental system (Fig. 3), they found liquid-like dynamics of shape relaxation in very good agreement with our experiments. Very recently, Bonazzi et al measured the viscosity in a related bacterial system, namely non-capsulated *Neisseria meningitidis* [25]. Using micropipette aspiration, they found a value of $\eta_{Nm siaD} = 50$ Nsm$^{-2}$, somewhat lower than our value. Remarkably, colonies formed by capsulated (wt) *N. meningitidis* have considerably lower viscosity and higher surface tension compared to *N. gonorrhoeae* studied here. Together with our results on systematic variations of motor properties this shows that fine-tuning T4P dynamics and binding kinetics has a tremendous effect on the materials properties of the colonies. In summary, the fusion dynamics of gonococcal colonies are in good agreement with the dynamics of shape changes expected during fusion of liquid drops whereby surface tension and viscosity are determined by the motor properties of the pili.

*Conclusion.* - We have shown that motor activity of T4P accelerates the processes of local ordering and shape relaxation during colony fusion. Our bacterial mutants allowed us to tune the frequency of T4P retractions and the detachment times between T4P of adjacent cells. This is most consistent with a role of T4P in increasing the attachment / detachment dynamics between bacteria accelerating the "equilibration" of the colony structure. For DNA-directed assembly, strong interactions tended to support fractal structures or small aggregates reminiscent of our force-deficient *ΔpilT* aggregates [1] (FIG. 4). We suggest that in our system cellular reproduction inhibits crystalline ordering because dumbbell formation before cell division introduces polydispersity and asymmetric cell shapes. Taken together, motor activity of T4P accelerates dense packing of bacteria, possibly allowing bacteria to protect themselves when facing external stress including antibiotic treatment or the prohibitive action of probiotic bacteria.

*Acknowledgements.* - We thank Nadzeya Kouzel, Katja Henseler, and Andrea Höne for experimental support, Katrina Forest and Michael Koomey for providing us with antibodies, the CECAD imaging facility of support with confocal microscopy, Kazem Edmont, Roel Dullens, Jan Dhont, and Stefan Egelhaaf for help with image analysis, Ramin Golestanian for helpful discussions, and the Deutsche Forschungsgemeinschaft and the Human Frontiers in Science Project for funding through grants MA3898 and RGP0061.

*Figure captions*

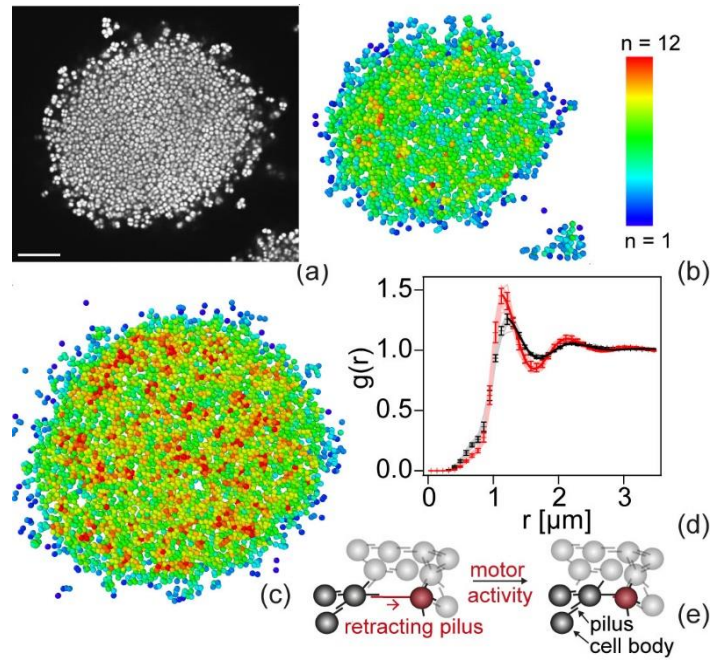

FIG. 1 (a) Confocal section of gonococcal wt* (Ng150) microcolony after 6 h. Reconstruction of 2 μm slices of bacterial coordinates after (b) 6 h and (c) 24 h of growth. The colors encode the number of nearest neighbors *n*. (d) Radial distribution functions *g(r)* after 6 h (black) and 24 h (red). Shaded lines: *g(r)* of individual colonies. Full line: fit to eq. 1. Dots and error bars: mean and standard deviation of 15 colonies from 3 independent experiments. Scale bar: 10 μm. (e) Hypothetical sketch of how T4P retraction supports local ordering.

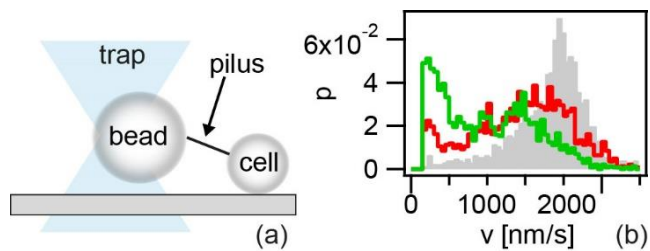

FIG. 2 Tuning T4P motor dynamics. (a) Experimental setup for characterizing the velocity of pilus retraction. (b) Velocity distribution of T4P retraction for force clamped at *F = 30 pN*. Grey: wt* (Ng170), red: *pilT$_{WB1}$* (Ng171), green: *pilT$_{WB2}$* (Ng176).

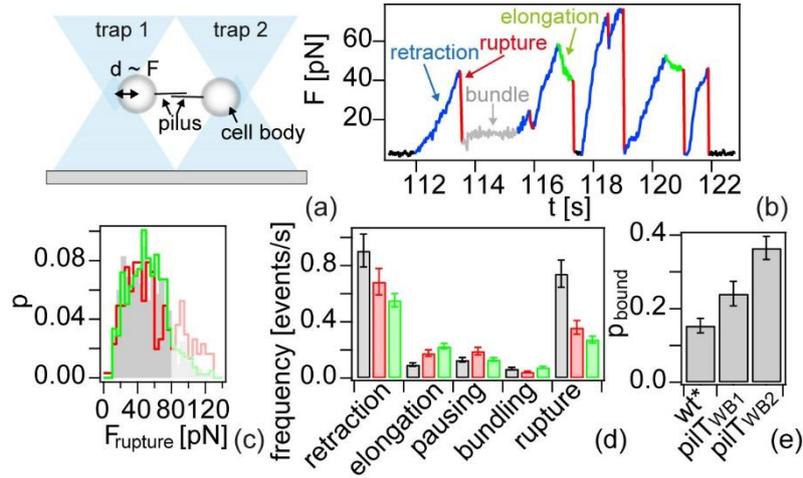

FIG. 3 Tuning bacterial interaction dynamics. (a) Experimental setup. A spherical cell is trapped in each of two laser traps. The deflection $d$ of the cell body from the center of the trap yields the force acting on both cell bodies $F \sim d$. When pili of both cells bind to each other and one of them retracts, both cell bodies are pulled towards each other. (b) Typical force $F$ as a function of time $t$. (c) Distribution of rupture forces for grey: wt* (Ng170), red: $pilT_{WB1}$ (Ng171), green: $pilT_{WB2}$ (Ng176). Forces were measurable up to 80 pN. (N > 300 for each strain) (d) Frequencies of T4P retraction elongation, pausing, bundling, and rupture. Color codes as in (c). (N = 38 - 844 per strain and condition, error bars: bootstrapping with N = 100) (e) Probability that at least one pair of pili from different cells are attached to each other. Error bars: bootstrapping.

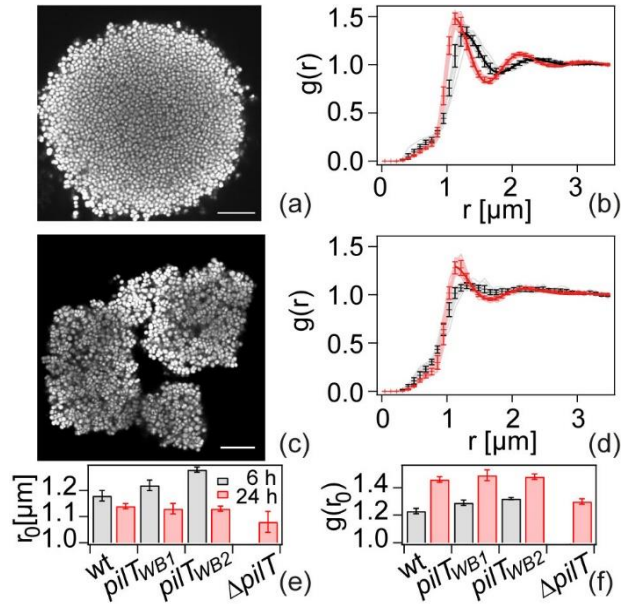

FIG. 4 (a) Typical confocal section after 6 h and (b) radial distribution functions $g(r)$ of T4P retraction reduced $pilT_{WB2}$ (Ng176) microcolony after 6 h (black) and 24 h (red). (c) Typical confocal section after 6 h and (d) radial distribution functions $g(r)$ of T4P retraction inhibited
11

*ΔpilT* (Ng178) microcolony after 6 h (black) and 24 h (red). Shaded lines: *g(r)* of individual colonies. Full line: fit to eq. 1. Dots and error bars: mean and standard deviation of at least 15 colonies from 3 independent experiments. Scale bar: 10 μm. (e) Contact distance $r_0$ and f) value of radial distribution function at $r_0$ obtained from fit to eq. 1.

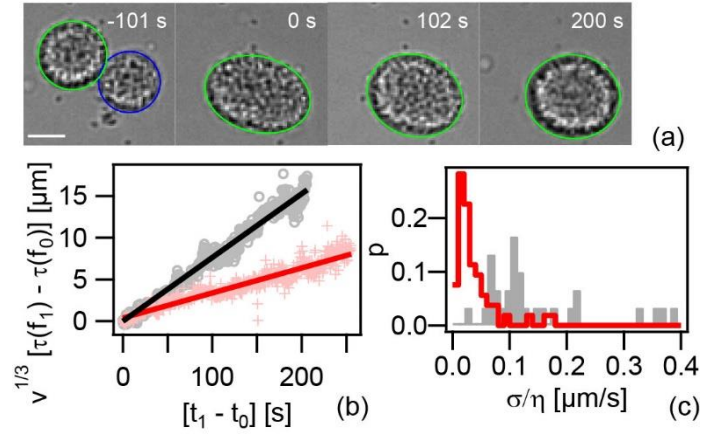

FIG. 5 (a) Time-lapse of wt* colony fusion event. (b) Typical $v^{1/3}[\tau(f) - \tau(f_0 = 0.71)]$ as a function of time. Grey circles: wt* (Ng151), red crosses: *pilT$_{WB1}$* (Ng171). Full lines: fits to eq. 3. (c) Distribution of ratio between surface tension $\sigma$ and viscosity $\eta$ obtained from eq. 3. Grey: wt* (Ng151), red: *pilT$_{WB1}$* (Ng171). N > 30 colonies.

# Supplementary Information

**Supplementary Materials and Methods**

*Laser Scanning Microscopy.* - After 6 or 24 h of growth, biofilms were stained with 10 µM Syto9 (Thermo Fisher Scientific) for 15 min at 37°C. Subsequently, the biofilms were fixed with 4% Formaldehyde in 1x PBS for 15 min at RT and washed with 1x PBS. For each experimental condition three separate experiments were performed and for each experiment five colony images acquired using a Leica TCS SP8 confocal laser scanning microscope (CECAD Imaging Facility) with a 63x, 1.4 NA, oil immersion objective lens. The excitation wave length was 488 nm. Image voxels were 35 nm x 35 nm x 130 nm.

The typical colony height is 60 µm and more. Using confocal microscopy, the image quality strongly depends on the location (Fig. S12a). Bacteria scatter and absorb the laser light of the excitation laser. We observed a tendency for lower fluorescence intensity of bacteria within the center of the colonies. This effect is likely to be caused by scattering of the laser beam at the bacteria between the bottom of the sample and the focal plane. Since the signal was high enough for determining the positions of the cells, we did not attempt to distinguish between these effects. Additionally, detector noise and PSF influence image quality. For quantification of the image quality, an example stack of a whole wt* colony was imaged after 24 h and the image signal and noise were determined separately. The image signal was defined with a mask around the tracked bacterial centroids and the noise was defined as the signal between the bacterial bodies. Up to 24 µm, signal and noise are clearly distinguishable (Fig. 12b). To ensure sufficient image quality and avoid boundary effects, we selected bacteria, which were located at a height of 8 to 22 µm and at least 1 µm away from the colony shell (Fig. 12c).

*Determination of bacterial coordinates.* - We used Matlab to analyse the structure of the raw biofilm image stacks. The particle tracking code was adapted from a version for 3D tracking [1] of the original IDL based code [2]. The code uses a spatial bandpass filter which smooths the image and subtracts the background off. As a result, we get smooth Gaussian-like blobs on a zero background. Next, the coordinates of the bacterial positions were found in pixel accuracy by searching for peaks in the filtered image. For a more precise estimate of the bacterial positions in sub-pixel accuracy, we put a mask around each bacterial position and calculated its centroid.

The goal is to determine parameters for the most accurate particle tracking. There are no optimal parameters for bacterial structures, because individual bacteria overlap (diplococci), Brownian motion of the fixed bacteria appears, the intensity of the detected signal decreases and noise increases with penetration depth into the bacterial structure. We set the Gaussian width of the bandpass filter to [1, 1, 1] pixel to obtain smooth Gaussian like blobs on zero background. The diameter of the bacteria was approximated with [0.7, 0.7, 0.7×2.1] µm. The factor 2.1 takes care of the broadened diffraction pattern in z-direction. The separation between the bacteria and the mask around the bacterial positions for the centroid calculation was set to 70% of the diameter. Please note that variation of these parameters does not significantly affect the positions of the peaks of the radial distribution function (Fig. S13).

Our image stacks consisted of 140 images. After filtering an image stack, we lost 30 images due to boundary effects. The remaining 110 images were analysed further. Particle positions, which are less than the diameter of a particle away from the edges were discarded. Especially, the bacteria on the colony shell (edge of colony) showed Brownian motion despite fixation. For that reason, we did not include bacteria, which were less than twice the bacterial diameter away from the colony shell.

We checked whether the radial distribution function was homogeneous within the analysed section. To assess cylindrical homogeneity, we subdivided the coordinate data into a cylinder with the major axis corresponding to the z-axis of the colony and a shell surrounding the central cylinder. We found no significant difference of the radial distribution functions (Fig. S14a, b). To assess vertical homogeneity, we subdivided the coordinate data into the lower half and the upper half of the analysed segment. The radial distribution functions were slightly shifted towards higher density in the upper half of the segment (Fig. S14c, d). Yet, the positions of the first maxima and their amplitudes were not significantly different, indicating that vertical inhomogeneity was negligible. Therefore, we did not consider spatial inhomogeneity within the colonies.

*Determination of radial distribution function (RDF).* - The radial distribution function is defined as the probability of finding two particles a distance $r$ apart, relative to the probability expected in an ideal gas at the same density. It was computed following Allen and Tildesley [3]. We determined the distances between all pairs of particles and sorted them into a histogram with $j$ bins and a width of $dr$. Assuming there are $n_{his}(b)$ pairs for a particular bin $b$ in the interval $[r, r + dr]$, the average number of particles whose distance from a given particle lies within the interval is: $n(b) = j$. The average number of particles in the same interval in an ideal gas is: $n_{id}(b) = \frac{4\pi v}{3}[(r+dr)^3 - r^3]$. The density $v$ was determined through a fit of $n_{id}(b)$ to $n(b)$. By definition the RDF is: $g(r + \frac{1}{2}*dr) = n(b)/n_{id}(b)$ [3].

We note that the shape of the RDF is influenced by the parameters set for obtaining the bacterial coordinates (Fig. S13). Variation of the width of the Gaussian filter barely influences the RDF (Fig. S13a). Varying the bacterial diameter shows minor effects at short distances (Fig. S13b). Since electron microscopy images are in good agreement with a bacterial diameter of 0.7 µm [4], this value was used in this study. The strongest effect was found, when the minimum separation between the bacteria was varied (Fig. S13c). In particular, for very small values, an additional peak around r = 0.5 µm occurred. We attribute this peak to the mean distance between the two cells of a diplococcus, i.e. a cell that has not finished separation and assumes the shape of a dumbbell. Here, we decided to use a minimal separation of 0.49 µm where the amplitude of the first maximum of g(r) was maximal. Importantly, the results show that the positions of the RDF peaks are independent of the parameter variation.

*Remoteness.* - The remoteness distribution function, a measure of the size of defects or inhomogeneity inside biofilm structures, gives the probability of finding a voxel inside the colony and a bacterial centroid a distance *r* apart. We implemented the remoteness distribution function [5], because the RDF averages over all particle distances and loses local information like defects in the structure. We generated a binary image by thresholding the original image to determine which voxels were inside the colony. Subsequently, we extruded 4.2 µm from the colony-edge and filled the holes inside the binary image. This function associates the foreground voxels of the binary image with the centroid of the closest bacterium. Note that the distance of the global maximum of the remoteness distribution is equal to the half distance of the global maximum of the RDF.

*Measurement of T4P retraction velocities.* - Retraction velocities were measured using an optical tweezers setup assembled on a Zeiss Axiovert 200 [6]. In short, all measurements were carried out in retraction assay medium consisting of phenol red-free Dulbecco's modified Eagle's medium (Gibco, Grand Island, NY) with 2 mM L-glutamine (Gibco), 8 mM sodium pyruvate (Gibco), 30mM HEPES (Roth) and 0 or 1 mM IPTG (Roth). A Suspension of Bacteria and carboxylated latex beads with a diameter of 2 µm (Polysciences, Warrington, PA) was applied to a microscope slide and sealed. All measurements were performed at 37 °C. The trap stiffness was determined by power spectrum analysis of the Brownian motion of a trapped bead to be 0.5 pN/nm ± 10%. The retraction velocities were measured in force clamp mode. During the experiment, a bead was trapped and held close to an immobilized bacterium at the surface. Eventually, a pilus attached to the bead and its retraction lead to a deflection out of the equilibrium position. As soon as the deflection of the bead reached the threshold deflection corresponding to a force of 30 pN, a force feedback algorithm held the displacement constant by moving the sample in the xy-plane using a piezo stage.

*Measurement of bacterial interaction forces.* - In order to characterize single cell interactions, an optical trap was installed in an inversed microscope (Nikon TI-E) [7]. The trapping laser (20I-BL-106C, Spectra Physics, wave length 1064nm, 5.4W) was coupled into the fluorescence port and focused into the sample through a water-immersion objective (Nikon Plan Apochromate VC 60x N.A. 1.20). Water immersion was used to prevent spherical aberrations and keep the trap stiffness independent of the distance to the glass surface. The position of the laser trap was controlled with a two-axis acousto-optical deflector (DTD-274HD6 Collinear Deflector, IntraAction Corp., USA). A square wave frequency signal with a frequency change rate of 100 kHz was applied to one axis of the acousto-optical deflector to create two spatially separated traps. The distance between the centers of the two traps was adjusted to be 2.84 µm, corresponding to 1.5 MHz in our setup. The position of bacteria in each trap was detected with brightfield images sequences recorded with a CCD camera (sensicam qe, PCO, Kelheim, Germany) at minimal ROI and maximum framerate. For every measurement the average framerate was recorded. The position of each bacterium for every point in time was calculated using the Hough transformation and MATLAB. The optical trap was calibrated by the equipartition theorem assuming a harmonic trap potential. The average stiffness over ten bacterial pairs was determined to be 0.11 (± 0.01) pN/nm. The optical restoring force increased linearly with the deflection of bacteria from the center of the trap up to deflections of 350 nm. Thus the forces are measureable up to 2 x 40 pN. In order to minimize stress responses due to laser induced heat, sample temperatures were kept at 33°C. The samples were prepared by selection of single piliated colonies from overnight GC-plates. The colonies were diluted in GC medium plus Isovitalex. Ascorbic acid (500 mM) was added to inhibit damage by radicals. Coverslips were etched with potassium hydroxide, coated with BSA and washed rigorously prior to usage. Measurements were performed with at least 5 µm distance to the coated glass surface in order to prevent cell-surface interaction. Only monococcal interactions were recorded. We did not detect any decrease of T4P - activity during 280 s. On average, bacteria were trapped for 69 s.

*Analysis of colony fusion dynamics.* - A custom build flow chamber was mounted into an inverted microscope (Nikon TI). The chamber was assembled around a Teflon holder with a single channel (length: 56 mm, width: 6 mm, height: 1 mm) by adding coverslips to the top and the bottom. The bottom coverslip was coated with BSA to prevent strong adhesion of the bacteria. Twinsil (Picodent) was used to fix the coverslips to the holder. An inlet and outlet were used to exchange the medium and inoculate bacteria. Bacteria were resuspended in GC containing 1% Isovitalex and diluted to an OD of 0.25. For *pilT$_{WB1}$*, IPTG was added to a final concentration of 1 mM.

Experimentally, wt* bacteria were inoculated without supply of oxygen. Within a period of ~ 30 min, bacteria used up the oxygen and as a consequence the colonies disassembled as described before [8]. Subsequently, oxygen-rich medium was applied triggering the formation of motile colonies (Fig. 5a). The initial step of oxygen-dependent disassembly was introduced to ensure that we analysed only colonies whose aggregation was reversible. We found that microcolony formation was reversible for wt* and *pilT$_{WB1}$* but not for the *pilT$_{WB2}$* strain, i.e. microcolonies formed by the latter strain did not disassemble when oxygen was depleted similar to *pilT* deletion strains [8].

Images were analyzed using Matlab and the DIPimage toolbox. For the analysis, colonies with a volume v < $10^4$ μm$^3$ were taken into account. First, the bright halo around larger objects as colonies was removed. Then the image was median filtered and the local variance was calculated. A threshold was applied to create a binary image and the holes of enclosed areas were filled to detect single objects. Binary operations were used to neglect the deformation of the detected colony contour originating from single and immobile bacteria on the coverslip in the close surroundings. The contour of the object was extracted and fitted by an ellipse for every single time point.

Gordon et al [9,10] proposed a model that yields the ratio between surface tension $\sigma$ and viscosity $\eta$ from the dynamics of the shape relaxation from an ellipsoid towards a sphere. Therein, the rate of rounding up of an ellipse is

$$\frac{df}{dt} = \frac{\sigma}{v^{1/3}\eta}\rho(f)$$

where $f = b/a$ is the ratio of the minor axis $b$ and major axis $a$ and the volume $4/3 \pi a b^2$, and

$$\rho(f) = \frac{3}{8(f^2-1)}\left(\frac{4\pi f}{3}\right)^{1/3}\left\{-2 - f^2 + \frac{f^2(4-f^2)}{(1-f^2)^{1/2}} \ln\left[\frac{1+(1-f^2)^{1/2}}{f}\right]\right\}$$

The expression for the time it takes for an ellipse to change from $f_0$ to $f_1$

$$\frac{\sigma}{\eta}[t_1 - t_0] = v^{1/3}\int_{f_0}^{f_1}\frac{df}{\rho(f)} = v^{1/3}[\tau(f_1) - \tau(f_0)]$$

was integrated numerically with $f_0 = 0.71$. The lower limit $f_0$ was chosen such that the aggregates were in good agreement with the shape of an ellipse. The ratio between surface tension $\sigma$ and viscosity $\eta$ was calculated by a linear fit.

Furthermore, we analysed shape relaxations during early time points following an approach proposed by Flenner et al. based on conservation laws proposed for the coalescence of highly viscous molten drops [11]. The fusing colonies are modeled as two spherical caps of radius $R(\Theta)$ (Fig. 9a) with a circular contact region of radius $r(\Theta) = R(\Theta)\sin\Theta$. Volume conservation requires

$$R(\theta) = 2^{\frac{2}{3}}(1+\cos\theta)^{-\frac{2}{3}}(2-\cos\theta)^{-\frac{1}{3}}R_0$$

with $R_0 = R(0)$. The rate of the decrease in surface energy is $\dot{W}_s = \sigma \, dS/dt$, where $S$ is the free surface area. The rate of energy dissipated by viscous flow is $\dot{W}_\eta \approx -4\pi R^4 \eta \alpha^2$, where $\alpha$ describes the biaxial stretching flow [11]. Using energy balance $\dot{W}_s = \dot{W}_\eta$ and assuming $R(\theta) \approx R_0$ they obtain

$$\left(\frac{r}{R_0}\right)^2 \approx A(t)\left[1 - e^{-\frac{t}{\tau_c}}\right] \quad (eq.\,4)$$

and

$$A(t) = 2^{\frac{4}{3}}\left(1 + e^{-\frac{t}{2\tau_c}}\right)^{-\frac{4}{3}}\left(2 - e^{-\frac{t}{2\tau_c}}\right)^{-\frac{2}{3}}$$

The characteristic time $\tau_c$ is related to the surface tension and the viscosity by $\tau_c = \eta R_0/\sigma$.

In our experiments, the circular contact region $r$ was described well by this model for longer time scales (Fig. S10b, c). For short time scales, the relaxation occurred faster than described by the model. When fitting the model (eq. 4) to the full data sets including all time points, the ratio of $\sigma/\eta$ was in agreement with the ratio derived from eq. 3 describing the relaxation at long time scales (Fig. S10d). We interpret the kinks (marked by arrows in Fig. S10b, c) as the time points of transition from early fast relaxation to slower liquid-like relaxation. On average these transitions occurred at (8 ± 3) s for the wt* strain and at (13 ± 7) s for strain *pilT$_{WB2}$*. These transition times were considerably shorter than the characteristic times obtained from the fits to eq. 4 of $\tau_{cwt*}$ = (63 ± 37) s and $\tau_{cpilTWB2}$ = (200 ± 110) s. We have shown previously, that gonococcal colonies form a fairly immobile core surrounded by a layer of highly motile cells [12], Video 2 therein. The thickness of the motile layer is on the order of ~ 2 bacterial diameters. We therefore propose that the mobile layer rapidly reorganizes once two colonies come into contact. As a consequence, the length of the neck region *r* increases faster than predicted by the fluid-model because it involves reorganization of bacteria residing directly at the surface independent of rearrangements within the bulk of the colony.

We conclude that shape relaxations of colonies are consistently described by two different models [9] [11] for fusion of liquid droplets at long time scales. Early fusion dynamics at a time scale of 10 s shows a different behavior consistent with a narrow layer of motile cells residing at the surface of the colonies.

*Estimation of surface tension and viscosity.* – Microscopically, we have determined the rupture force between T4P $F_{rupture} \approx 50$ pN and the probabilities that bacteria are connected to each other by T4P p$_{bound\ wt*} \approx 0.15$ and p$_{bound\ WB1} \approx 0.25$ (FIG. 3). These values enable us to estimate the surface tension $\sigma = F\Delta x\ /\Delta A$ where $F\Delta x$ is the work required to increase the surface area by $\Delta A$. The work required for moving one bacterium from the bulk to the surface by the distance *D* is estimated as follows. The diameter of a bacterium is D ≈ 0.7 µm, the average number of T4P per cell is N ≈ 7 [4], and the increase in surface area is $\Delta A \approx \pi(D/2)^2 \approx 0.4$ µm$^2$. We assume that half of the pili are not bound when the bacterium resides at the surface. Together, we estimate

$$\sigma = \frac{1}{2} N p_{bound} \frac{F_{rupture}\ D}{\pi \left(\frac{D}{2}\right)^2}$$

we thus obtain $\sigma_{wt*} \approx 5 \cdot 10^{-5}$ Nm$^{-1}$ and $\sigma_{WB1} \approx 8 \cdot 10^{-5}$ Nm$^{-1}$.

Using these values, our experimentally determined ratios of $\sigma / \eta$ (FIG. 5) allow estimating the viscosities. We obtain average values of $\eta_{wt*} \approx 350$ Nsm$^{-2}$ and $\eta_{WB1} \approx 2000$ Nsm$^{-2}$.

*Experimental estimate of surface tension.* We verified experimentally, that the surface tension estimated above provides the right order of magnitude. To this end, microcolonies were deformed by centrifugation and the Young-Laplace equation was used to estimate the surface tension [13]. Approximating the shape of the deformed colony by an ellipsoid [14], the surface tension can be determined by measuring the radius of the microcolony (top view in the microscope) before and after centrifugation through

$$\sigma = \Delta \rho g^* r_{ell}^2 \, G$$

where $\Delta \rho$ is the density difference between the surrounding medium and the cells of the aggregate, $g^*$ is the centrifugation force in relative gravitational units, $r_{ell}$ is the main semi-axis of the ellipse, and $G$ is a geometric factor given by $G = (a^9 + a^3 - 2)^{-1}$, with $a = r_{ell}/r_{sphere}$, the latter being the radius of the spherical aggregate prior to centrifugation. As such, the surface tension can be obtained by the simple knowledge of $r_{ell}, r_{sphere}$.

wt* (Ng151) were grown overnight on GC Agar (see below). 5-10 colonies were resuspended in 500 μL fresh medium, vortexed for 5 minutes, and placed in an ibidi 8-well μ-slide (Ibitreat). Prior to centrifugation, the $r_{sphere}$ of the colonies was determined using an inverted confocal microscope (Nikon C1 confocal microscope, 60x magnification) which scans the full height of the colony in slices of 1 μm (30 slices) and a typical lateral resolution of 256 x 256 pixel for a field of view of around 30 x 30 μm (FIG. S11a). Subsequently, samples were centrifuged at 200 g for 1 - 5 min. Aggregates were then fixed with a 4% paraformaldehyde solution after centrifugation to avoid shape relaxation and imaged again (FIG. S11b). Subsequent image analysis was performed in ImageJ. Here, the main axis $r_{ell}, r_{sphere}$ are determined using a plot profile of the intensity profile at the largest width of the colony.

The buoyant density of *E. coli* was determined to be in the range of (100 - 200) fg [15]. Assuming the same density and a volume of 0.5 μm³, $\Delta \rho \sim$ (200 - 400) kg m⁻³. Using these parameters, we obtain σ ≈ 1.5 · 10⁻⁵ N/m, i.e. at the same order of magnitude compared to our estimate.

*Growth conditions.* - Gonococcal base agar was made from 10 g/l BactoTM agar (BD Biosciences, Bedford, MA, USA), 5 g/l NaCl (Roth, Darmstadt, Germany), 4 g/l K2HPO4 (Roth), 1 g/l KH2PO4 (Roth), 15 g/l BactoTM Proteose Peptone No. 3 (BD), 0.5 g/l soluble starch (Sigma-Aldrich, St. Louis, MO, USA)) and supplemented with 1% IsoVitaleX: 1 g/l D-Glucose (Roth), 0.1 g/l L-glutamine (Roth), 0.289 g/l L-cysteine-HCL×H20 (Roth), 1 mg/l thiamine pyrophosphate (Sigma-Aldrich), 0.2 mg/l Fe(NO3)3 (Sigma-Aldrich), 0.03 mg/l thiamine HCl (Roth), 0.13 mg/l 4-aminobenzoic acid (Sigma-Aldrich), 2.5 mg/l β-nicotinamide adenine dinucleotide (Roth) and 0.1 mg/l vitamin B12 (Sigma-Aldrich). GC medium is identical to the base agar composition, but lacks agar and starch.

*Bacterial biofilm growth conditions.* - Each biofilm was grown within ibidi μ-Slides I^0.8 Luer flow chambers for 6 or 24 h, respectively, at constant nutrient flow of 3 ml/h by using a peristaltic pump (model 205U; Watson Marlow, Falmouth, United Kingdom). Bacteria from overnight plates of each strain were resuspended in GC medium to an optical density at 600nm (OD600) of 0.01 and 300 μl of cell suspension was inoculated into the flow chambers. The bacteria were left for 1 h at 37°C to allow for attachment to the glass surface. After 1 h, the flow was switched on.

*Bacterial strains.* - All bacterial strains were derived from the gonococcal strain MS11 (VD300). In all strains, we deleted the G4-motif by replacing it with the *aac* gene conferring resistance against apramycin. The G4-motif is essential for antigenic variation of the major pilin subunit [16]. Pilin antigenic variation modifies the primary sequence of the pilin gene. Since the composition of amino acids can affect the pilus density and potentially the rupture force between pili, antigenic variation is likely to generate heterogeneity and distract form the major topic of this study. Since there is no reason to expect that deletion of the G4 motif affects the structure of the biofilm, the strain carrying the G4 deletion is labeled wt* throughout the manuscript. To support this assumption, we verified that the radial distribution function *g(r)* for the unmodified wt VD300 was comparable to *g(r)* for wt* shown in Fig. 1 (Fig. S4). Furthermore, strains expressing *gfpmut3* and *mcherry*, respectively, were used for some of the experiments. Since these genetic modifications do not affect growth rate or biofilm structure [17] and carried the same *pilE* sequence, they were labeled wt* in the text as well. All figure captions denote the exact strains used for the specific experiments.

We generated strains *pilT$_{WB1}$* and *pilT$_{WB2}$* that produced a non-functional form of the T4P retraction protein PilT. Since PilT forms hexameric rings [18], we hypothesized that non-functional PilT had a dominant negative effect of T4P retraction dynamics. For tuning the dynamics, two strains were generated whereby non-functional *pilT$_{WB}$* was expressed from promoters with different strengths. Here we were interested in the phenotypic effects of PilT$_{WB}$ production, i.e. the fact that the T4P retraction frequency and the rupture frequency were affected (Fig. 3). Therefore, the relative levels of functional and non-functional PilT were not characterized.

PilT is an ATPase whose activity depends on the Walker A box for ATP binding and on the Walker B box for ATP hydrolysis. Exchanging the glutamate within the Walker B box by alanine results in non-functional PilT$_{WB}$. To this end, *pilT* was amplified by PCR, using primers to replace E206A, resulting in *pilT$_{WB}$*. For generating strain *pilT$_{WB1}$* with *pilT$_{WB}$* expression under the control of the IPTG-inducible *lac* promoter (Ng171), the PCR fragment was restricted using PacI and FseI and cloned into the pGCC4 integration vector designed for integration between *lctP* and *aspC*. pGCC4 was a gift from Hank Seifert (Addgene plasmid # 37058). *P$_{lac}$pilT$_{WB}$* was integrated by transforming Ng150 with pGCC4-*P$_{lac}$pilT$_{WB}$* by standard transformation protocols. This strain was induced by adding 1mM IPTG to the agar plates and the growth medium.

For generating strain *pilT$_{WB2}$* with *pilT$_{WB}$* expression under the control of the strong *pilE* promoter (Ng176), the sequence of the PCR fragment encoding for *pilT$_{WB}$* was fused to a PCR fragment amplified from genomic DNA containing the *pilE* promoter. The fragment was cloned into the SacI site of the integration vector pIGA [17] generating *pIGA-P$_{pilE}$pilT$_{WB}$*. *P$_{pilE}$pilT$_{WB}$* was integrated by transforming Ng150 with *pIGA-P$_{pilE}$pilT$_{WB}$* by standard transformation protocols.

For strain *pilT$_{WB2}$*, we verified that *pilT$_{WB}$* was expressed as follows. We generated strain *pilT$_{WB2}$ pilT$_{his}$* (Ng158) expressing His-tagged PilT in the native *pilT* locus and *pilT$_{WB}$* under control of the *pilE* promoter in the *igA1* locus. The His-tag was introduced by transformation with a PCR-fragment containing the C-terminal sequence of *pilT*, the sequence encoding for six histidines, and a sequence downstream of *pilT*. This strain was transformed with *pIGA-P$_{pilE}$pilT$_{WB}$*. Subsequently, we performed a Western blot with an antibody against PilT. The His-tag allowed separating the bands formed by PilT-His and by PilT$_{WB}$. Ng158 whole cell protein extracts were prepared by sonifying cell material of a complete grown agar plate for 3 min. After adding Triton X-100 to a final concentration of 0.5%, samples were incubated for 30 min on ice. After centrifugation for 30 min at 16000 RCF the protein concentration was determined by Bradford. Proteins were separated by SDS–PAGE in TG-prime gels (SERVA) containing 14% acrylamide, transferred onto nitrocellulose membranes using wet electrotransfer. Membranes were blocked with 5% milk in TBST (10 mM Tris, 150 mM NaCl, 0.1% Tween-20) and incubated in the specific antiserum (anti-PilT at 1/3000, anti-GC at 1/2000) followed by horseradish peroxidase (HRP)-coupled anti-rabbit antibody (1/10 000). Membranes were developed by enhanced chemiluminescence ECL-plus (GE-Healthcare) and recorded using the Lumiimager. By quantifying the two bands generated by PilT-His and PilT$_{WB}$, respectively (Fig. S5a), we found that the protein ratio was PilT$_{WB}$ : PilT-His = 2.1 ± 0.4. The error is the standard deviation from three replicates. Note that in strain Ng158 *recA* was inducible (to inhibit pilin antigenic variation [19] [16]). T4P dynamics are comparable in strains expressing *recA* [20] and not expressing *recA* [21]. *pilT* is expressed both in the presence [6] and in the absence [22] of *recA*. Therefore, the lack of *recA* expression was not relevant for showing that *pilT$_{WB}$* is expressed.

For strain *pilT$_{WB1}$*, we verified that *pilT$_{WB}$* was expressed as follows. Strain *pilT$_{WB1}$* was transformed with genomic DNA of the *ΔpilT* strain (Ng178), generating *pilT$_{WB1}$ ΔpilT* (Ng182). Ng182 whole cell protein extracts were prepared as follows. Cells were harvested from agar plates and used to inoculate 5 ml liquid cultures with a starting OD600 of 0.05, supplemented with 1 mM or without IPTG. After overnight incubation (37 °C, 5% CO2, 250 rpm shaking) cells were pelleted by brief centrifugation, subsequently resuspended in lysis buffer (Tris-HCL pH 8.5, 2 mM EDTA, 0.1% Triton-X) and boiled for 5 minutes at 95 °C. After centrifugation (30 min at 16000 RCF, 4 °C) the protein concentration was determined by Bradford. SDS-PAGE and Western blotting were performed as described above. In the absence of induction of *pilT$_{WB}$*, no band was visible (Fig. S5b). When Ng182 was induced with 1mM IPTG, the *pilT$_{WB}$* band was visible showing that *pilT$_{WB}$* was expressed.

Deletion of the retraction ATPase *pilT* abolishes the ability of T4P to retract and to generate force [23]. The ability to form pilus polymers remains intact [22]. The *pilT* deletion strain (Ng178) was generated by transforming Ng150 with chromosomal DNA from strain GT17 [24] and selecting for resistance against chloramphenicol.

Correct insertions were verified by sequencing.

*Bacterial two-hybrid assay.* - Chemically competent BTH101 cells (Table S1) were co-transformed with combinations of the different BACTH plasmids containing the ORFs of *pilT* and *pilT$_{WB}$*, respectively (Table S3). Serial dilutions of the transformants were plated on LB plates supplemented with /X-gal, IPTG, 100 μg/ml Ampicillin and 50 μg/ml Kanamycin plates and incubated at 30°C for 48 hours in order to obtain about 100-200 colonies per plate. If the proteins interact with each other the resulting β-galactosidase activity results in blue colonies. As a positive control, competent cells were co-transformed with the control plasmids pKT25-zip and pUT18C-zip whereas in the negative control cells were co-transformed with the plasmids pKT25 or pKNT25 and pUT18 or pUT18C. Bacterial cells to be assayed for ß-galactosidase activity were grown in 5 ml of LB broth in the presence of 0,5mM IPTG and appropriate antibiotics at 30°C overnight. 1.5 ml of the liquid cultures were centrifuged at 17.8 rcf for 1 min, resuspended in 1 ml 0,85% NaCl and OD600 was recorded. 0.5ml cells were mixed with 0.5ml buffer Z (40 mM $Na_2HPO_4 \cdot 7H_2O$, 60 mM $NaH_2PO_4 \cdot H_2O$, 10 mM KCl, 1 mM $MgSO_4 \cdot 7H_2O$, 50 mM 2-ß-Mercaptoethanol). To lyse the cells, 20 μl chloroform and 10 μl 0,1% SDS were added and each sample were vortexed. Samples were incubated 10 min at 30°C. To initiate the reaction, 200 μl ONPG was added to each sample and the samples were incubated at 30°C. Until a yellow color developed, the reaction was stopped by adding 0,5 ml 1M $Na_2CO_3$. After incubation time of 5 min the OD420 and OD550 were determined of each sample. The unit for this assay is equal to micromoles of ONPG hydrolyzed per minute per OD600 unit.

**Supplementary Figures**

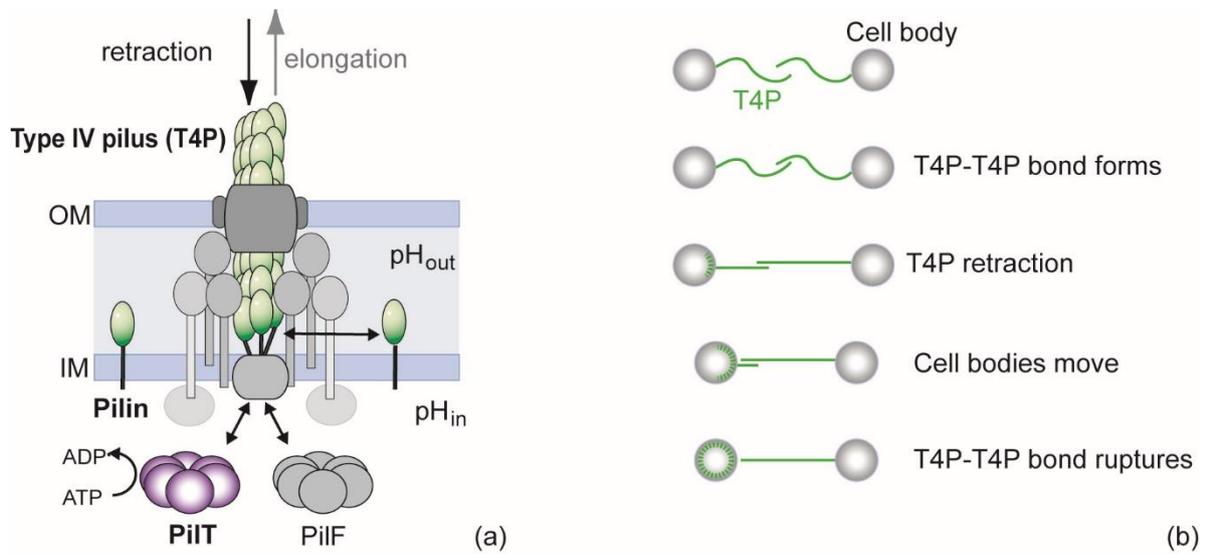

FIG. S1 Model of type IV pilus (T4P) motor dynamics and force generation. (a) The T4P is a polymeric filament formed by pilin subunits. Pilins are stored in the inner membrane (IM) and can be integrated into the elongating pilus. Elongation of T4P requires the hexameric elongation ATPase PilF. The hexameric pilus retraction ATPase PilT is essential for pilus retraction. We consider the entire T4P complex as a molecular motor generating force by T4P retraction. The chemical energy powering pilus retraction is most likely provided through ATP hydrolysis [18] by PilT and by proton motive force [25]. (b) T4P-T4P binding and retraction govern bacterial interactions. T4P of adjacent cells bind to each other. When one of the pili retracts by depolymerization, then force is generated between the cells, shortening their relative distance.

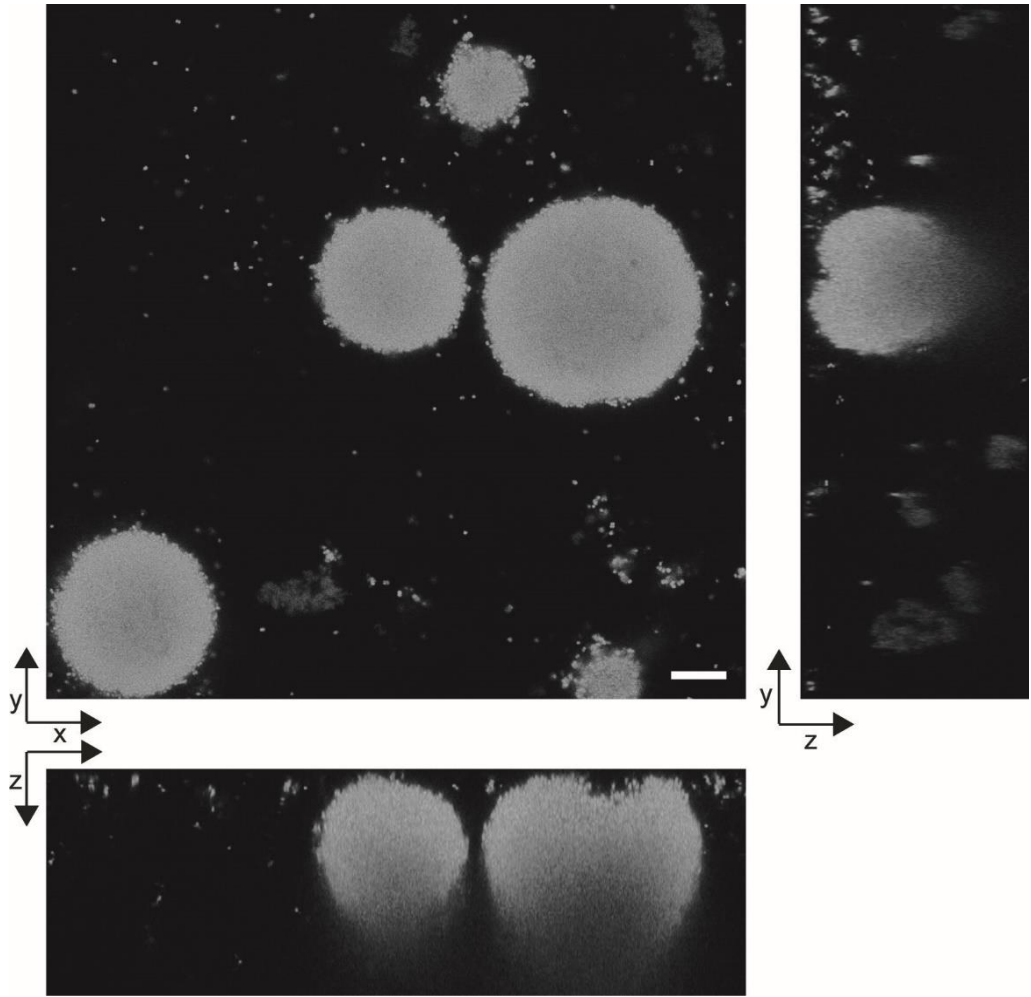

FIG. S2 Typical biofilm structure after 24 h (Ng150) of growth in flow chamber. Orthogonal view. Scale bar (x, y, z): 20 μm.

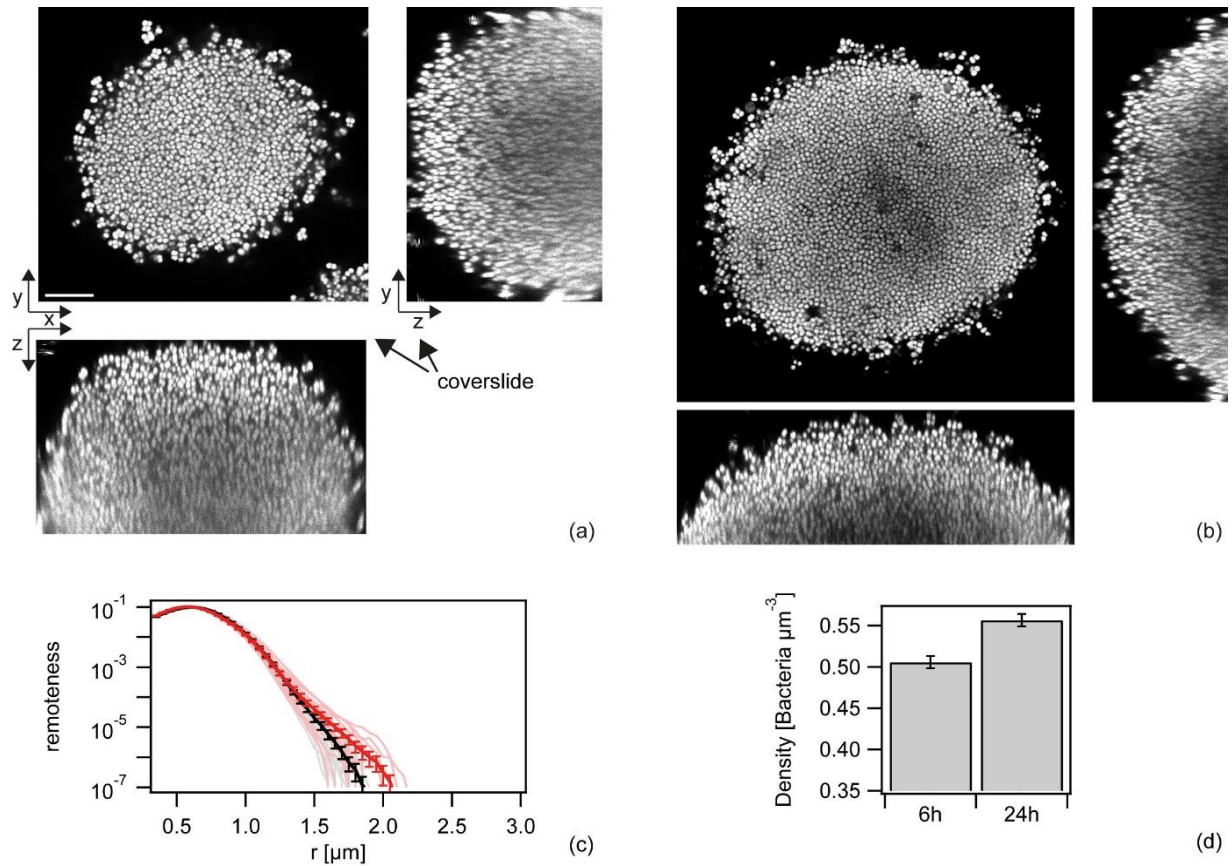

FIG. S3 Typical orthogonal views of wt* (*ΔG4*, Ng150) colonies formed after (a) 6 h and (b) 24 h. (c) Remoteness distribution after 6 h (black) and 24 h (red). (d) Number of bacteria per volume. Error bars: standard errors of 15 colonies from three independent experiments.

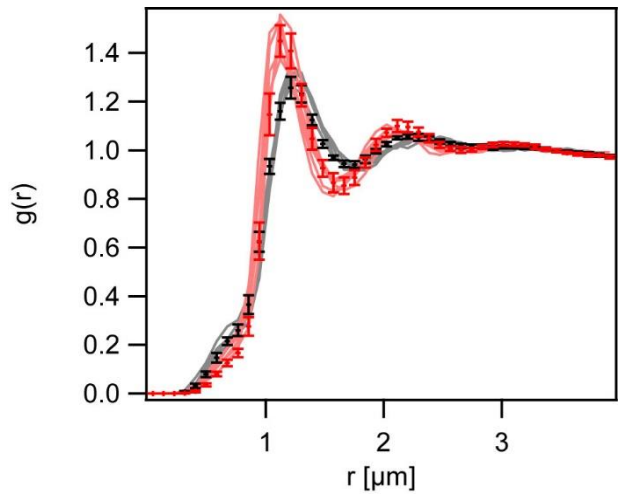

FIG. S4 Radial distribution function *g(r)* of wt gonococci (VD300). Shaded lines: *g(r)* of individual colonies after 6 h (black) and 24 h (red). Dots and error bars: Mean and standard deviation of wt* as shown in Fig. 1d.

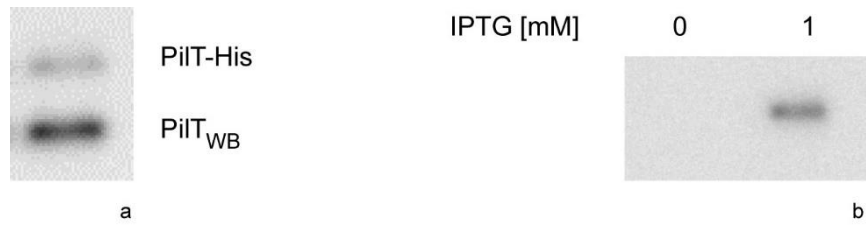

FIG. S5 Verification of *pilT$_{WB}$* expression. a) Control for production of PilT$_{WB}$ in strain *pilT$_{WB2}$*. Western blot of strain *pilT$_{WB2}$ pilT$_{his}$* (Ng158). Due to fusion of native PilT with the His-tag, the band separates from the PilT$_{WB}$ band. b) Control for production of PilT$_{WB}$ in strain *pilT$_{WB1}$*. Western blot of strain *pilT$_{WB1}$ pilT* (Ng182). Due to deletion of native *pilT*, only PilT$_{WB}$ was produced. Since *pilT$_{WB}$* was under the control of an IPTG-inducible promoter, PilT$_{WB}$ was generated only in the presence of IPTG.

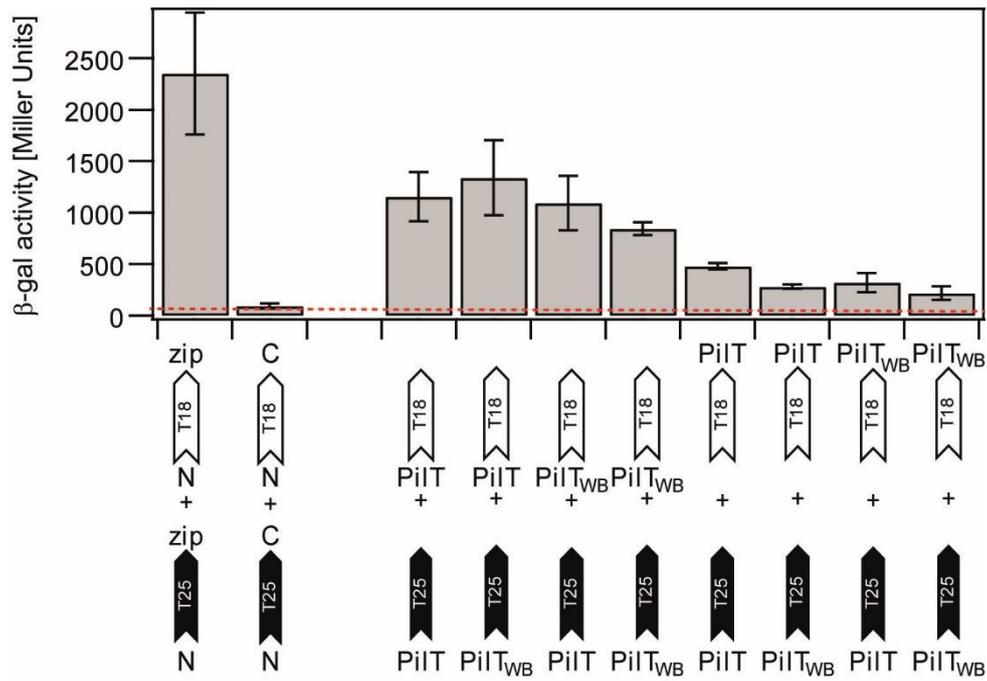

FIG. S6 Bacterial two-hybrid assay characterizing the interaction between PilT and PilT$_{WB}$. First column: positive control, second column: negative control. Protein interactions were quantitatively determined by measuring β-galactosidase activities from *E. coli* BTH101 cells co-transformed with the indicated proteins fused to the adenylate cyclase T25 or T18 subunit. The β-galactosidase activities are expressed in Miller Units as mean values of three individually selected replicate co-transformants, each measured in triplicates. Error bars: standard deviation from three independent experiments.

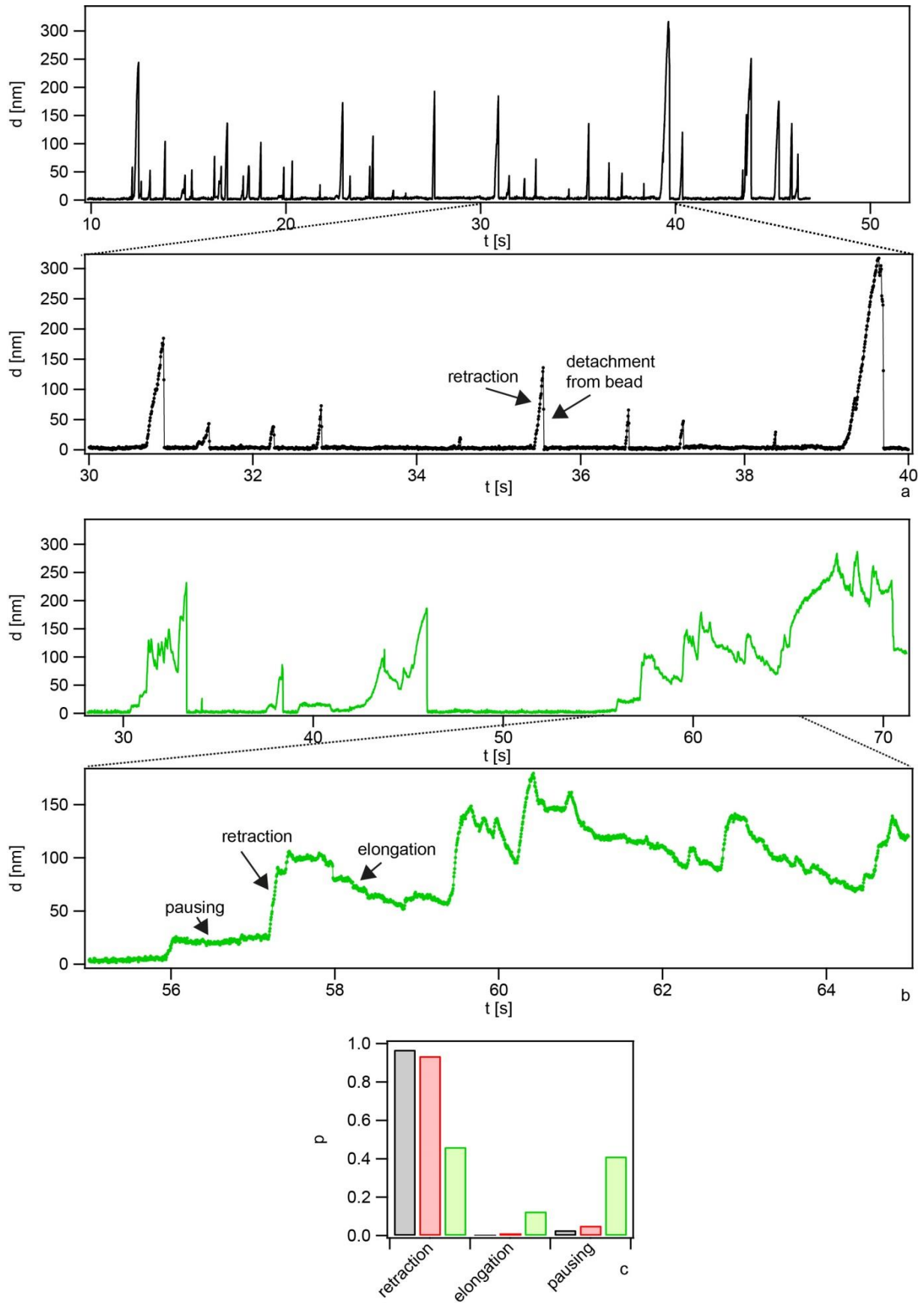

Fig. S7 Time distribution of occupation of retractions-, elongation, and pausing state at F = 30 pN determined from T4P retraction assay (Fig. 2a). a) Typical event observed with wt*

(Ng170). b) Typical event observed with *pilT$_{WB2}$* (Ng176). c) Probability of finding a T4P in the states of retraction, elongation, or pausing. Grey: wt* (Ng170), red: *pilT$_{WB1}$* (Ng171), green: *pilT$_{WB2}$* (Ng176). N > 1400 events for each condition.

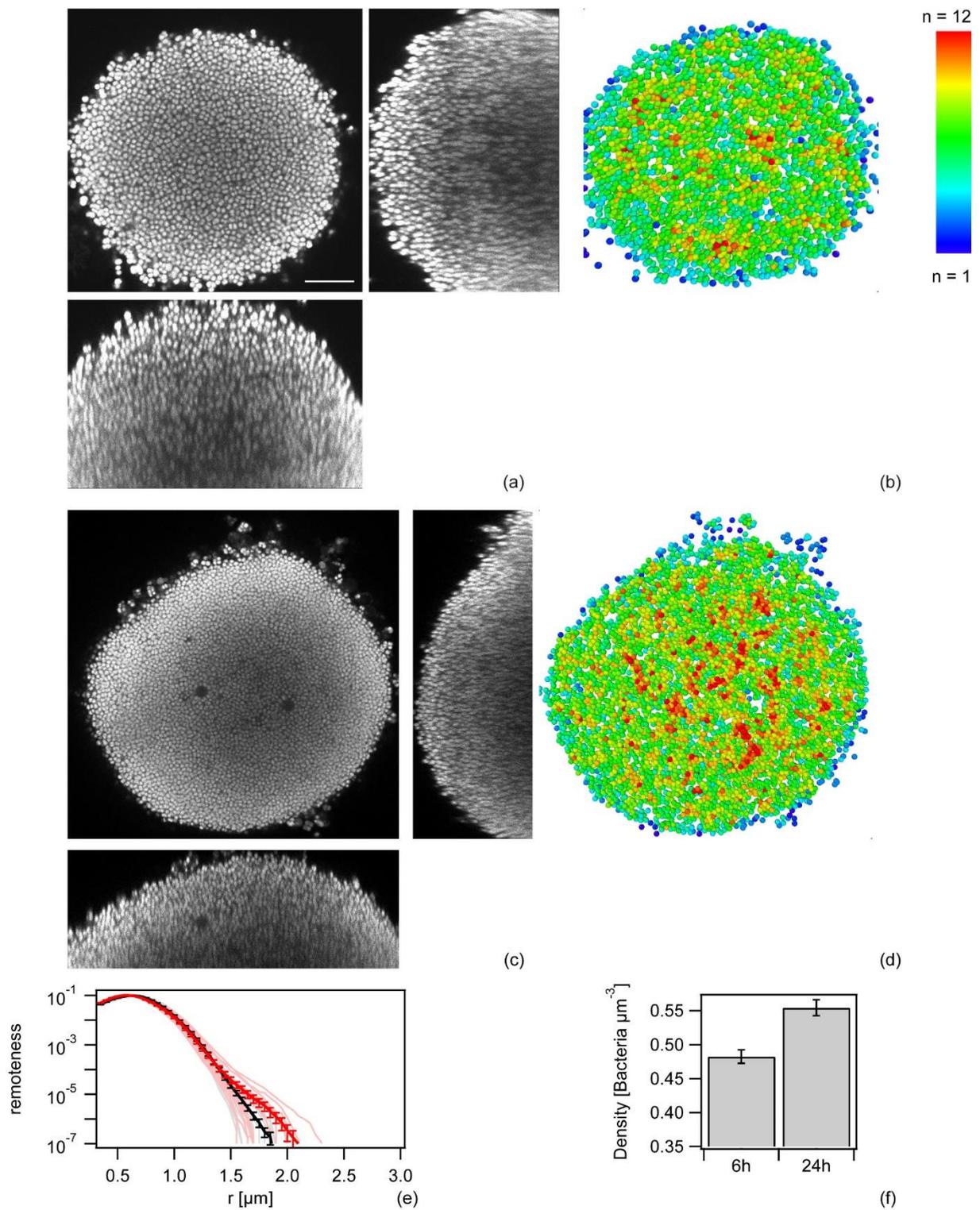

FIG. S8 Typical colonies of *pilT$_{WB2}$ ΔG4* (Ng176) colonies formed after (a), (b) 6 h and (c), (d) 24 h. (a), (c) Orthogonal views. (b), (d) Reconstruction of bacterial coordinates. The colors encode the number of nearest neighbors *n*. (e) Remoteness distribution after 6 h (black) and 24 h (red). (f) Number of bacteria per volume. Error bars: standard errors of 15 colonies from three independent experiments. Scale bar 10 μm.

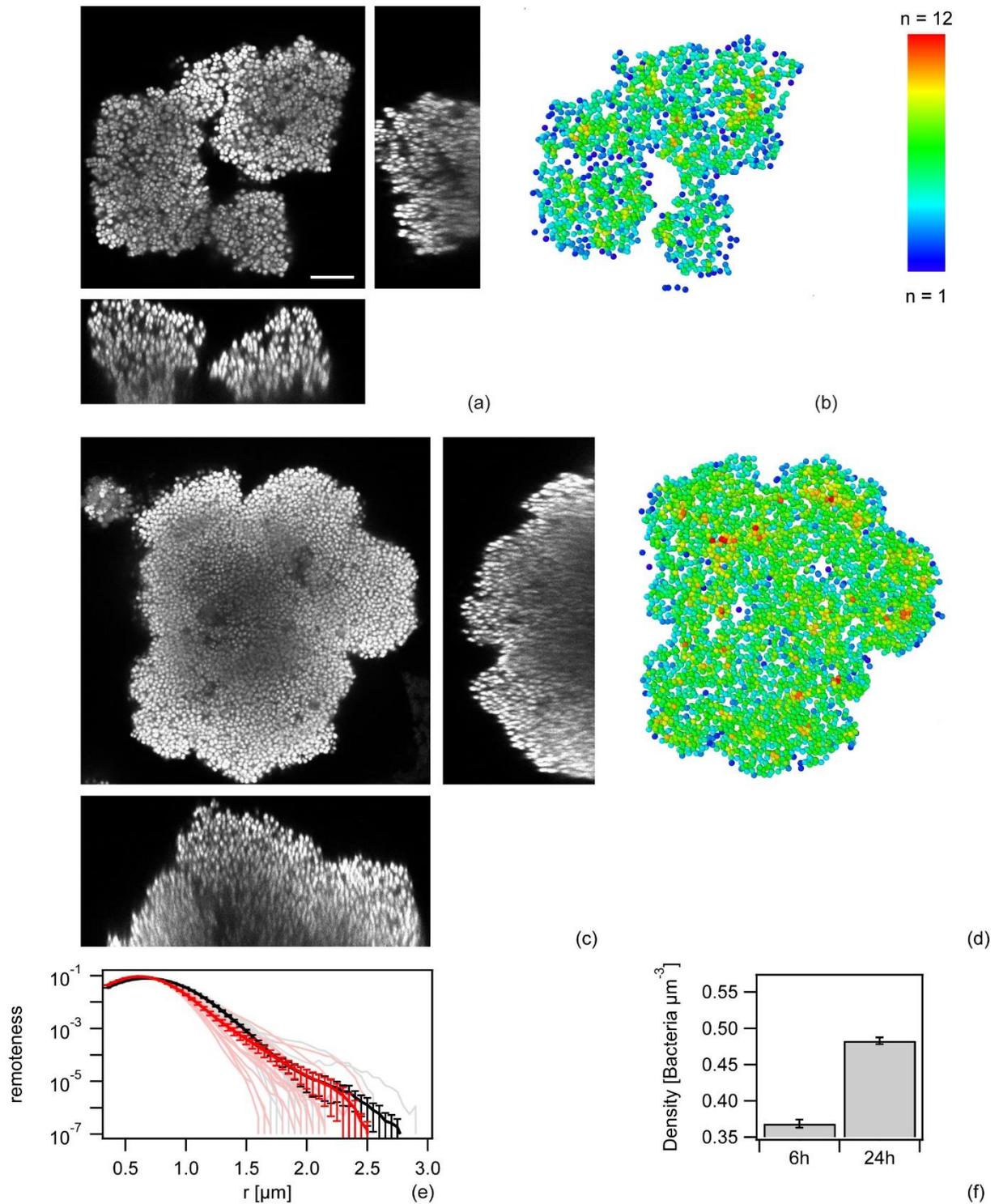

FIG. S9 Typical colonies of *ΔpilT ΔG4* (Ng178) colonies formed after (a), (b) 6 h and (c), (d) 24 h. (a), (c) Orthogonal views. (b), (d) Reconstruction of bacterial coordinates. The colors encode the number of nearest neighbors *n*. (e) Remoteness distribution after 6 h (black) and 24 h (red). (f) Number of bacteria per volume. Error bars: standard errors of at least 15 colonies from three independent experiments. Scale bar 10 μm.

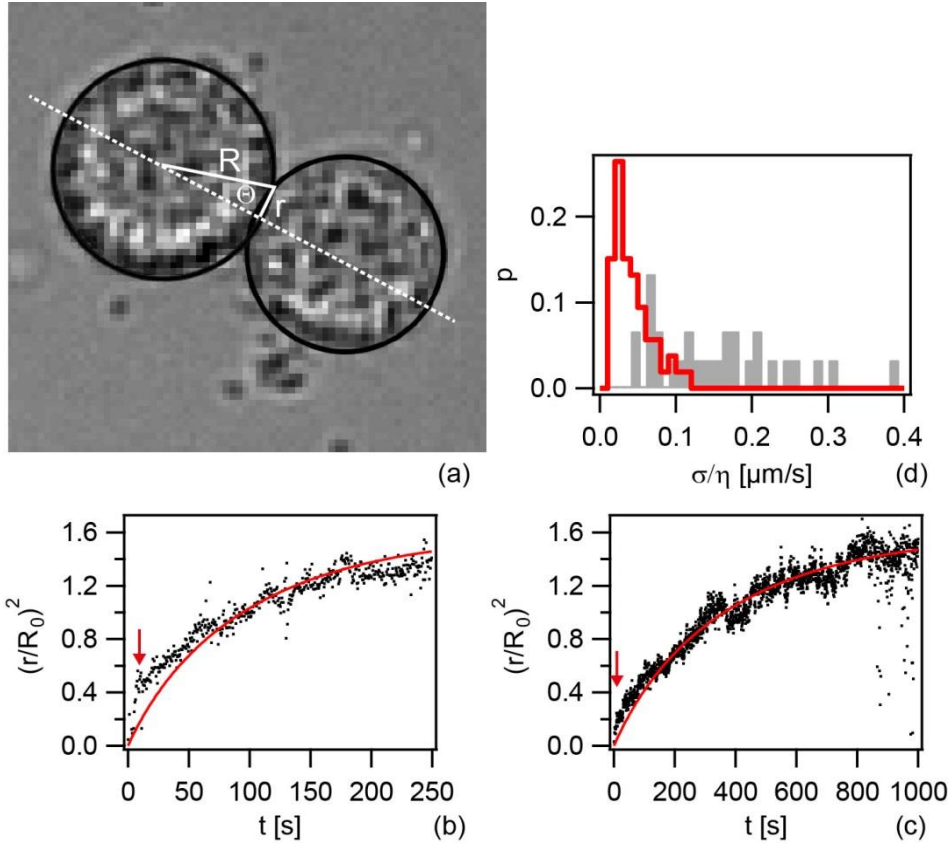

FIG. S10 Early shape relaxations during colony fusion deviate from fluid-like behavior. (a) Example of two fusing colonies. They are described as two spherical caps with radius $R$ and circular contact region $r$. Circular contact region $r$ normalized to initial cap radius $R_0$ as a function of time for (b) wt* (Ng151) and (c) $pilT_{WB1}$ (Ng171). Full red lines: Fits to eq. 4. Red arrows: transition from early regime of fast relaxation to slow fluid-like relaxation. (d) Distribution of ratio between surface tension $\sigma$ and viscosity $\eta$ obtained from eq. 4. Grey: wt* (Ng151), red: $pilT_{WB1}$ (Ng171). $N > 30$ colonies.

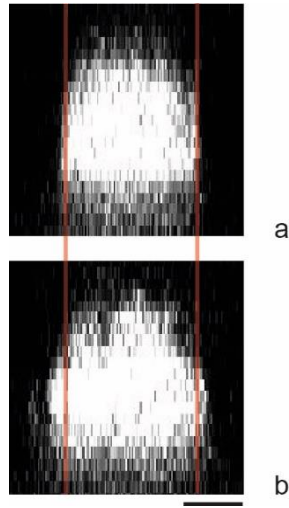

FIG. S11 Experimental test of surface tension. x-z images of wt* (Ng151) colonies were image a) prior to and b) after centrifugation at 200 g.

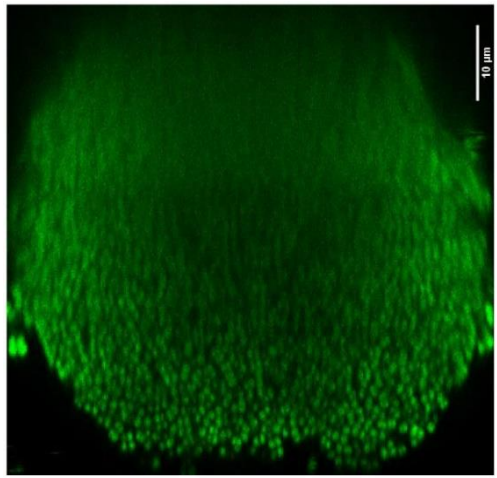
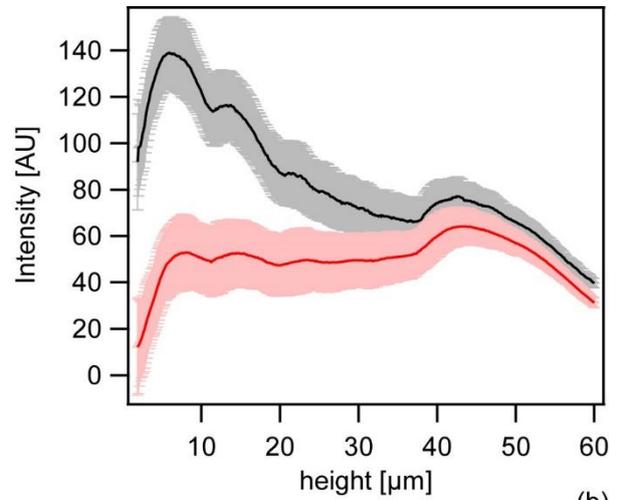
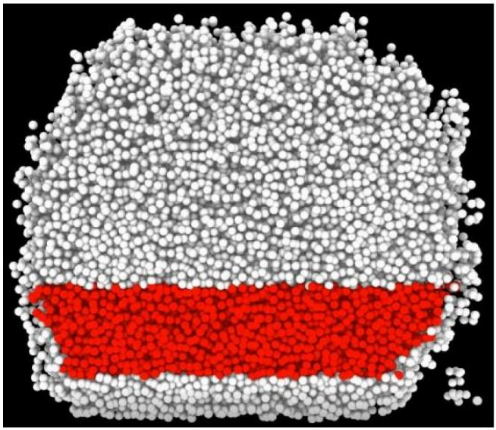

FIG. S12 Determination of analysed segment. (a) Cross section of an example stack of a whole wt* colony after 24 h. Scale bar: 10 µm (b) Mean and standard deviation of signal and noise. Signal and noise were determined separately for each layer from the example stack. The image signal was defined with a mask around the tracked bacterial centroids and the noise was defined as the signal between the bacterial bodies. (c) Tracked bacterial position from example stack are illustrated with white spheres (diameter = 0,7 µm). Selected bacterial positions for analysis are color coded in red.

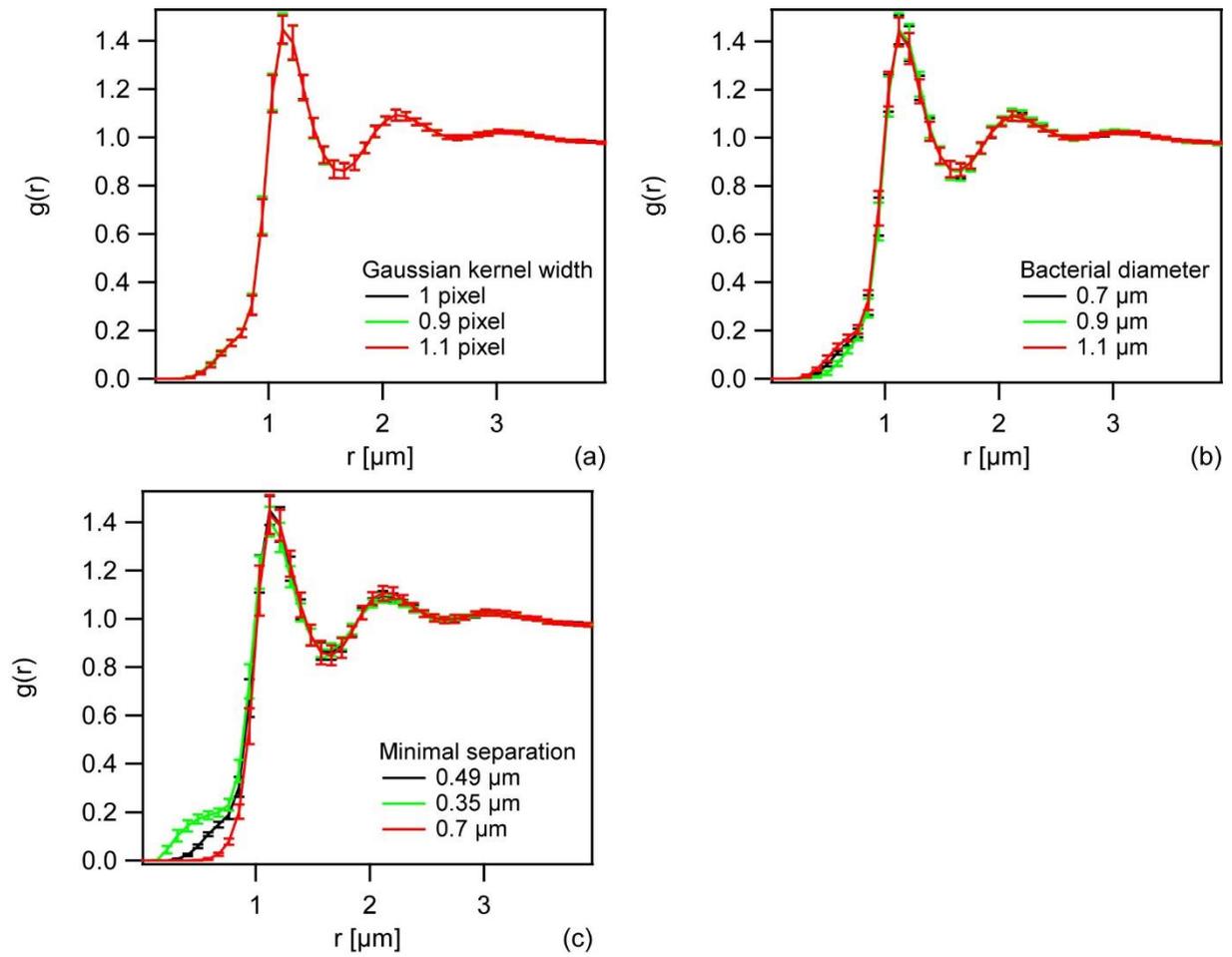

FIG. S13 Effect of parameter variation for coordinate determination on radial distribution function $g(r)$. wt* (Ng150) colonies were analyzed after 6 h; error bars: standard deviation of 15 colonies from 3 independent experiments. $g(r)$ for varying (a) width of the Gaussian filter, (b) bacterial diameter, and (c) separation between the bacteria and the mask around the bacterial positions.

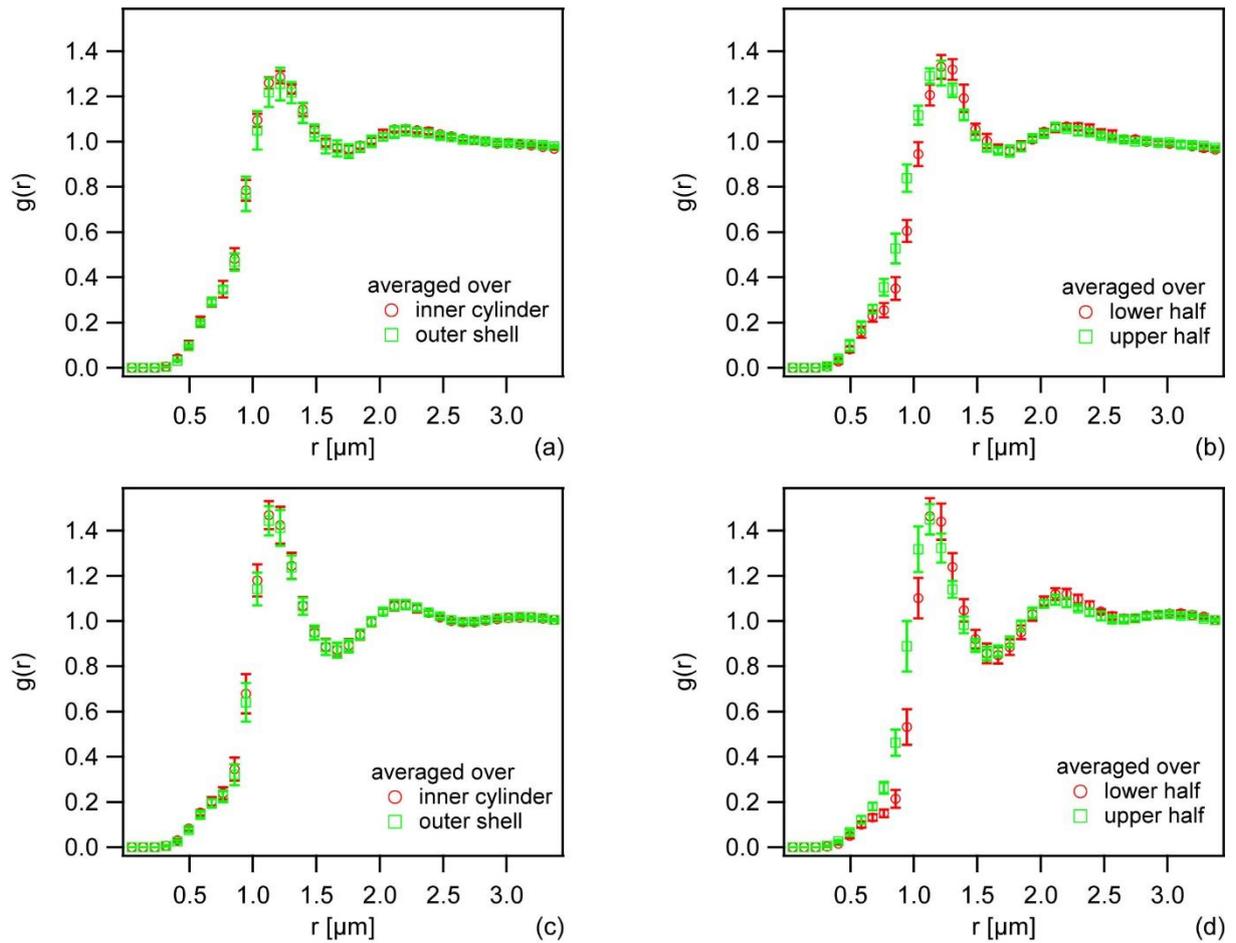

FIG. S14 Evaluation of radial and vertical homogeneity within colony. Radial distribution functions *g(r)* of wt* (Ng150) colonies were analyzed after 6 h (a, c) and 24 h (b, d), respectively. a, b) Data from z-stack containing 110 images were split into a cylinder around the central axis of the colony (red circles) and a shell surrounding the central cylinder (green squares). c, d) Data from z-stack containing 110 images were split into the lower 55 images (red circles) and the upper 55 images (green squares). Error bars: standard deviation of 15 colonies from 3 independent experiments.

**Supplementary Tables**

| Strain | Relevant genotype | Source/Reference |
|---|---|---|
| *wt* VD300 (Ng002) | wild type, *opa*- selected | [26] |
| *wt\** (Ng150) | *G4::aac* | [27] |
| *wt\** (Ng151) | *iga::P$_{pilE}$ gfpmut3 ermC* <br> *G4::aac* | This study |
| *wt\** (Ng170) | *lctp:mcherry aadA:aspC* <br> *G4::aac* | [27] |
| *pilT$_{WB1}$* (Ng171) | *lctp:P$_{lacP}$ pilTWB ermC:aspC* <br> *G4::aac* | This study |
| *pilT$_{WB2}$* (Ng176) | *iga::P$_{pilE}$ pilTWB ermC* <br> *G4::aac* | This study |
| *pilT$_{WB2}$ pilT$_{his}$* (Ng158) | *pilT-his* <br> *iga::P$_{pilE}$ pilTWB ermC* <br> *P$_{lac}$ recA tetM* | This study <br> [19] |
| *pilT* (Ng178) | *pilT::m-Tn3cm* <br> *G4::aac* | [24], This study |
| *pilT$_{WB1}$ pilT* (Ng182) | *lctp:P$_{lacP}$ pilTWB ermC:aspC* <br> *pilT::m-Tn3cm* <br> *G4::aac* | [24], This study |
| BTH101 (*E. coli*) | F$^-$, *cya-99, araD139, galE15, galK16, rpsL1 (Str$^R$), hsdR2, mcrA1, mcrB1* | EUROMEDEX |

Table S1 Strains used in this study

|  | $r_0$ | $m$ | $g(r_0)$ | $l$ | $a$ | $b$ |
|---|---|---|---|---|---|---|
| wt* 6h | 1.1762 ± 0.0249 | 2.5384 ± 0.612 | 1.2694 ± 0.0161 | 0.16254 ± 0.0356 | 3.03 ± 0.296 | -6.5198 ± 0.393 |
| wt* 24h | 1.1365 ± 0.0116 | 1.9641 ± 0.598 | 1.457 ± 0.015 | 0.38182 ± 0.0246 | 2.4294 ± 0.139 | -6.6956 ± 0.202 |
| pilT$_{WB1}$ 6h | 1.2152 ± 0.0174 | 3.3962 ± 0.553 | 1.2882 ± 0.0185 | 0.20053 ± 0.0315 | 3.2044 ± 0.18 | -6.8177 ± 0.261 |
| pilT$_{WB1}$ 24h | 1.1309 ± 0.017 | 3.827 ± 0.657 | 1.4874 ± 0.0437 | 0.35792 ± 0.0411 | 2.3909 ± 0.109 | -6.8723 ± 0.221 |
| pilT$_{WB2}$ 6h | 1.2826 ± 0.0343 | 2.5445 ± 0.000641 | 1.3179 ± 0.0282 | 0.23231 ± 0.0203 | 3.2922 ± 0.246 | -7.1311 ± 0.254 |
| pilT$_{WB2}$ 24h | 1.1311 ± 0.01 | 1.975 ± 0.615 | 1.4809 ± 0.015 | 0.41214 ± 0.0225 | 2.2951 ± 0.113 | -6.7632 ± 0.173 |
| ΔpilT 6h |  |  |  |  |  |  |
| ΔpilT 24h | 1.0823 ± 0.0381 | 2.4398 ± 0.602 | 1.2962 ± 0.0218 | 0.13771 ± 0.0608 | 2.6537 ± 0.384 | -5.5219 ± 0.492 |

Table S2 Fit parameters for radial distribution functions

| Plasmid | | |
|---|---|---|
| pKNT25 | Bacterial Andenylate Cyclase Two-Hybrid System vector; Kan$^R$ | EUROMEDEX |
| pUT18 | Bacterial Andenylate Cyclase Two-Hybrid System vector; Amp$^R$ | EUROMEDEX |
| pKT25 | Bacterial Andenylate Cyclase Two-Hybrid System vector; Kan$^R$ | EUROMEDEX |
| pUT18C | Bacterial Andenylate Cyclase Two-Hybrid System vector; Amp$^R$ | EUROMEDEX |
| pKT25-zip | Bacterial Andenylate Cyclase Two-Hybrid System vector; Kan$^R$ | EUROMEDEX |
| pUT18C-zip | Bacterial Andenylate Cyclase Two-Hybrid System vector; Amp$^R$ | EUROMEDEX |
| pKKS10 | Plasmid used for BACTH analysis. Contains PCR product encoding *pilT*, cloned in KpnI and PstI site of pKNT25; Kan$^R$ | |
| pKKS11 | Plasmid used for BACTH analysis. Contains PCR product encoding *pilT*, cloned in KpnI and PstI site of pUT18; Amp$^R$ | |
| pKKS12 | Plasmid used for BACTH analysis. Contains PCR product encoding *pilT*, cloned in KpnI and PstI site of pKT25; Kan$^R$ | |
| pKKS13 | Plasmid used for BACTH analysis. Contains PCR product encoding *pilT*, cloned in KpnI and PstI site of pUT18C; Amp$^R$ | |
| pKKS15 | Plasmid used for BACTH analysis. Contains PCR product encoding *pilT$_{WB}$*, cloned in KpnI and PstI site of pKNT25; Kan$^R$ | |
| pKKS16 | Plasmid used for BACTH analysis. Contains PCR product encoding *pilT$_{WB}$*, cloned in KpnI and PstI site of pUT18; Amp$^R$ | |
| pKKS17 | Plasmid used for BACTH analysis. Contains PCR product encoding *pilT$_{WB}$*, cloned in KpnI and PstI site of pKT25; Kan$^R$ | |
| pKKS18 | Plasmid used for BACTH analysis. Contains PCR product encoding *pilT$_{WB}$*, cloned in KpnI and PstI site of pUT18C; Amp$^R$ | |

Table S3 Plasmids used for the Bacterial Two Hybrid Assay.